\documentclass[a4paper]{jpconf}
\usepackage{amsmath}
\usepackage{amsfonts}
\usepackage{amssymb}
\usepackage{anyfontsize}
\begin{document}
\title{On special limits of the Mixed Painlev\'e P$_{\mathbf{III-V}}$ Model}

\author{V C C Alves${}^1$, H Aratyn${}^2$, J F Gomes${}^1$,  
and A H Zimerman${}^1$}

\address{${}^1$ S\~ao Paulo State University, UNESP\\
Instituto de F\'\i sica Te\' orica - IFT/UNESP\\
Rua Dr. Bento Teobaldo Ferraz, 271\\
01140-070, S\~ ao Paulo - SP, Brazil}
\address{${}^2$ Department of Physics \\
University of Illinois at Chicago\\
845 W. Taylor St.\\
Chicago, Illinois 60607-7059\\
}

\ead{victorcca17@gmail.com, aratyn@uic.edu, francisco.gomes@unesp.br, zimerman@ift.unesp.br}

\begin{abstract}
The paper discusses P$_{III-V}$ equation for special values of its
parameters for which this equation reduces to  P$_{III}$, I$_{12}$, as well as,
to some special cases of I$_{38}$ and I$_{49}$  equations
from the Ince's list of $50$ second order differential equations possessing 
Painlev\'e property.

These reductions also  yield symmetries governing the 
reduced models obtained from the P$_{III-V}$ equation.
We point out that the solvable equations on 
Ince's list emerge in this reduction scheme when the underlying 
reflections of the Weyl symmetry group no longer include
an affine reflection through the hyperplane orthogonal to the
highest root and therefore do not give rise to an affine Weyl group.
We hypothesize that on the level of the underlying algebra and geometry
this might be a fundamental feature that 
distinguishes the  six Painlev\'e
equations from the remaining $44$ solvable equations on the Ince's list.

\end{abstract}


\section{ Introduction }
Painlev\'e equations emerged in a study of ordinary second order
differential equations with solutions that have no movable critical
points other than poles. Equations with such characteristic referred
to as Painlev\'e Property \cite{conte} can be identified with one
of 50 canonical types listed by Ince \cite{Ince}. Forty four  of these equations 
can be either linearized or are solvable in terms of known 
transcendental functions. The relevant, for this paper, examples are equations 
I$_{12}$, I$_{38}$ and I$_{49}$ listed in 
\ref{sec:ince}. 
The  remaining six equations
are known as Painlev\'e P $_I$, P $_{II} $,\ldots,P$_{VI}$ equations, see
\ref{sec:ince} 
for explicit expressions of equations P$_{III}$ and P$_{V}$.

One of the most fundamental developments in the study of integrable models
has been Ablowitz, Ramani and Segur \cite{ars} conjecture
that partial differential evolution equations of integrable 
hierarchies reduce 
in self similarity limit to differential equations 
with Painlev\'e Property.
In particular the 2M-Boson integrable model \cite{2mboson} obtained 
as reductions of KP integrable models connected to Toda lattice 
hierarchy gives rise to Painlev\'e equations invariant under extended 
affine Weyl groups.
It was shown in reference \cite{AGZ2016} that the 4-Boson integrable 
model (M=2), can be reduced after elimination of a pair of degrees of freedom by 
Dirac reduction in a self-similarity limit to a  mixed $P_{III-V}$
equation, namely, 
\newpage
\begin{align} 
q_{zz} & = - \frac1z q_z+ \left( \frac{1}{2q}+\frac{1}{2(q-r_1)} \right)
\left(q_z^2 -\epsilon_0^2 r_0^2 z^{-2-2J}\right)
- (2 \alpha_2+\alpha_1+\alpha_3-1) \frac{(q- r_1) q r_0}{z} \nonumber \\
&+\frac{r_0^2}{2} q (q-r_1) (2q-r_1)
+\frac{\alpha_1^2 r_1 (q-r_1)}{2 z^2 q}-
\frac{\alpha_3^2 r_1 q}{2 z^2 (q-r_1)}  \label{qzz}\\
&+ \frac{\epsilon_0r_0 z^{-J-2} }{q (q-r_1)} \Big\lbrack 
(\alpha_1+\alpha_3-J)q^2 +
qr_1 (-2 \alpha_1+J)+\alpha_1 r_1^2\Big\rbrack
-2 \epsilon_1 r_1 z^{J-1} q (q-r_1) \, .
\nonumber
\end{align}

This equation fulfills the necessary condition for having the
Painlev\'e Property and  further reduces to P$_{III}$ and 
P$_{V}$ equations for special values of its parameters.
Here $J, \epsilon_0,\epsilon_1,r_0,r_1$ together with  $ \alpha_j, j=0,1,2,3$ (with
$ \sum_{j=0}^3 \alpha_j =1$)
define the extended parameter space of mixed P$_{III-V}$ model.

In this paper we systematically study submodels and their symmetries that are
obtained from P$_{III-V}$ model for special values of its parameters. 
For the purpose of this study it is convenient to 
alternatively define P$_{III-V}$ equation
in terms of  symmetric equations:
\begin{align}
z f_{i,\,z} &=f_i  f_{i+2} \big(
 f_{i+1}- f_{i+3} 
 \big)+(-1)^{i} f_i \,\big(\alpha_{1}+\alpha_{3} -(1+J)/2 \big)+
\alpha_i\big(f_{i}+f_{i+2}\big)  \label{big}
\\ 
&-(-1)^{[i/2]} \epsilon_{i+1} \big(f_{i+1}+f_{i+3}\big), \qquad i=0,1,2,3
\nonumber
\end{align}
for $\epsilon_0=\epsilon_2, \epsilon_1=\epsilon_3$ and with 
symbol $[i/2]$ that is $i/2$, if $i$ is even or $(i + 1)/2$, if $i$ is odd.

For equations (\ref{big}) the constraints:
\begin{equation}
f_1+f_3 =r_1 z^{(1+J)/2}, \quad
f_0+f_2 = r_0 z^{(1-J)/2 } \, ,
\label{condcbig}
\end{equation}
are automatically satisfied with $r_0,r_1$ being integration constants.
Equations (\ref{big}) are obtained 
when the P$_{V}$ Hamiltonian  (see e.g. \cite{masuda,mok,noumiy}):
\begin{align}
h_0 &= f_0 f_1 f_2 f_3+\frac14 (\alpha_1+2\alpha_2-\alpha_3) f_0 f_1
+ \frac14 (\alpha_1+2\alpha_2+3 \alpha_3) f_1 f_2 \nonumber \\
&-\frac14 (3\alpha_1+2\alpha_2+\alpha_3) f_2 f_3 
+\frac14 (\alpha_1-2\alpha_2-\alpha_3) f_0 f_3+ 
\frac14 (\alpha_1+\alpha_3)^2 \nonumber
\end{align}
is augmented by two symmetry breaking terms: 
\begin{equation}
{\bar h}_0= h_0 +\frac{\epsilon_0}{2} (f_0^2-f_2^2)
+\frac{\epsilon_1}{2} (f_1^2-f_3^2).
\label{hbarzero}
\end{equation}
These terms break the $A^{(1)}_3$ symmetry  of P$_{V}$  equation down to 
invariance under one single automorphism operation : 
\begin{align}
 \pi(\alpha_i)&=\alpha_{i+1}, \qquad  \pi(f_i)=f_{i+1}, \qquad
 i=0,1,2,3\label{ext-pi1}\\
 \pi (\epsilon_0)&= \epsilon_1,  \qquad \pi (\epsilon_1)=- \epsilon_0,
 \quad \pi(J)=-J, \qquad \pi (r_0) =r_1, \qquad \pi (r_1)=r_0\, ,
\label{ext-pi2}
\end{align}
such that $\pi^4=1$. 
Defining canonical variables $q,p$ as:
\begin{equation}
q= f_1\, z^{-(1+J)/2}, \; p=-f_2 \, z^{(1+J)/2},
\label{qpf}
\end{equation}
one finds that equation (\ref{big}) is equivalent to the two first-order Hamilton equations:
\begin{align}
z q_z &= q \left(q-r_1 \right) \left(2p+r_0 z \right)- 
\left(\alpha_1+\alpha_3 \right) q
+\alpha_1 r_1  +\epsilon_0 r_0 z^{-J} \nonumber \\
z p_z &=   p \left(p+ r_0 z \right)
\left(r_1 - 2q \right) + 
(\alpha_1+\alpha_3) p - 
\alpha_2  r_0 z -
\epsilon_1 r_1 z^{J+1}
\label{qzpz}
\end{align}
that lead back to P$_{III-V}$ equation (\ref{qzz}) upon elimination of $p$.
Equations (\ref{qzpz}) follow from the Hamiltonian:
\begin{equation}
zH = q \left(q-r_1 \right) p \left(p+ r_0 z \right)-
\left(\alpha_1+\alpha_3 \right) q p
+\left( \alpha_1 r_1 +\epsilon_0 r_0 z^{-J}\right)p
+\left( \alpha_2  r_0 z+\epsilon_1 r_1 z^{J+1}\right)q\, ,
\label{Hamiltonian}
\end{equation}
which agrees with the Hamiltonian (\ref{hbarzero}) up to a constant.
The above automorphism $\pi$ from relation (\ref{ext-pi1}) can be rewritten in terms of canonical
variables as 
\begin{equation}
\pi(q)=-p/z, \qquad \pi(p)=(q-r_1)z, \qquad \pi(\alpha_i)=
 \alpha_{i+1}, 
 \label{pipbarq}
\end{equation}
and $\pi$ as defined above and in relation (\ref{ext-pi2})  
keeps  equations (\ref{qzpz}) invariant.
 
 The P$_{III}$ or P$_{V}$ Painlev\'e models emerge from  P$_{III-V}$
 for different 
values of the underlying parameters.
See below the list $i) - v)$ \cite{AGZ2016,AAGZ2018} for 
a complete summary of
models that can be obtained from P$_{III-V}$, their symmetries and the corresponding values
of parameters. The notation $W [s_1,s_3,\pi^2]$ used below denotes the symmetry 
 group generated by $s_1,s_3,\pi^2$.
\begin{enumerate}
\item[i)] P$_{III-V}$  defined for 
$r_0 \ne 0$ and $r_1 \ne 0$, $J\ne 0$ 
is invariant under automorphism $\pi$ for $\epsilon_0 \ne 0$
and $\epsilon_1 \ne 0$.
\item[ii)] P$_{III-V}$  defined for 
$r_0 \ne 0$ and $r_1 \ne 0$, $J\ne 0$ with only  one of the 
parameters $\epsilon_0$ (or $\epsilon_1$) being $ \ne 0$ 
is invariant under the extended affine Weyl group $W [s_0,s_2,\pi^2]$ (
or $W [s_1,s_3,\pi^2]$). Note that $\pi^2$ remains a symmetry even with
one of the $\epsilon_i$ parameters being set to zero.
\item[iii)]   P$_{V}$ (see equation (\ref{P5})) is obtained for 
$r_0 \ne 0$ and $r_1 \ne 0$, and either $J=0$ or parameters $\epsilon_i
=0 $ for   $i=0,1$ and is invariant under the $A^{(1)}_3$ symmetry
$W [s_0,s_1,s_2,s_3,\pi]$. 
\item[iv)] P$_{III}$  (see equation (\ref{P3})) is obtained in a limit when
either $r_0 \to0$ and $J \ne -1$ or $r_1 \to0$ and $J\ne 1$ and is
invariant under the extended affine Weyl group $W [s_0,s_2,\pi_0,\pi_2,\pi^2]$ 
(or $W [s_1,s_3,\pi_1,\pi_3,\pi^2]$) . It is possible to realize this
symmetry as $C^{(1)}_2$ extended affine Weyl group \cite{AGZ2016}.
\item[v)] Ince's equations XII ($I_{12}$),  (incomplete) XXXVIII
($I_{38}$) and XLIX ($I_{49}$) are 
obtained as a limit when
either $r_0 \to0$ and $J =  -1$ or $r_1 \to0$ and $J= 1$. The symmetry
is still  $W [s_0,s_2,\pi_0,\pi_2,\pi^2]$ (or $W [s_1,s_3,\pi_1,\pi_3,\pi^2]$) 
but actions of $\pi_i$ on $\alpha_j$ become identical to those of
$s_i$  and consequently the $C^{(1)}_2$ realization can no longer be established.
\end{enumerate}

In the next two sub-sections we will  give more detailed discussion of 
limits $ r_i \to 0,\, i=0,1$ discussed in cases iv)
and v) with special attention to symmetries valid at these limits
for various values of the parameter $J$. 

\section{The $\mathbf{r_1 \to 0}$ limit of  P$_{\mathbf{III-V}}$ model} 
Setting $r_1 = 0$ in  (\ref{qzz}) yields
\begin{equation} 
q_{zz} = - \frac{q_z}{z}+ 
\frac{\left(q_z^2 -\epsilon_0^2 r_0^2 z^{-2-2J}\right)}{q}
- (2 \alpha_2+\alpha_1+\alpha_3-1) \frac{q^2 r_0}{z}+r_0^2 q^3 
+ \epsilon_0r_0 z^{-J-2}(\alpha_1+\alpha_3-J)
\label{qzzr10}
\end{equation}
For the special value of $J=-1$ this equation takes form of the conventional 
Painlev\'e III equation (\ref{P3}) \cite{okap3} invariant under 
$W[ s_0,s_2,\pi_0, \pi_2, \pi^2]$ \cite{AGZ2016}.

However for arbitrary  values of $J$ equation (\ref{qzzr10})  remains invariant 
under 
\begin{align}
\pi_0(q)&= -\frac{\epsilon_0 z^{-(1+J)}}{q}, \quad  
\pi_0 (p)= \frac{ z^{(1+J)}}{\epsilon_0 } \left( q^2p+\alpha_2 q \right)\nonumber \\
\pi_0 (\alpha_1+\alpha_3) &=  J+1-2\alpha_2-\alpha_1-\alpha_3, \; \;
\pi_0 (\alpha_2)=  \alpha_2 , \; \;
\pi_0(\alpha_0)= 1-J-\alpha_0
\label{pi0}
\end{align}
and
\begin{align}
\pi_2(q)&= \frac{\epsilon_0  z^{-1-J}}{q}, \quad  
\pi_2 (p)= -\frac{ z^{1+J}}{\epsilon_0 } \left( q^2(p+r_0z)+(1-\alpha_2
-\alpha_1-\alpha_3)q\right)-r_0 z \nonumber \\
\pi_2(\alpha_1+\alpha_3) &=  J-1+2\alpha_2+\alpha_1+\alpha_3, \; \;
\pi_2(\alpha_2) =  1-J - \alpha_2 , \; \;
\pi_2 (\alpha_0)= \alpha_0\, ,
\label{pi2}
\end{align}
which  formally generalize to all values of $J$ the transformations 
that kept P$_{III}$ invariant for $J=-1$ \cite{AGZ2016}.

In addition to (\ref{pi0}) and (\ref{pi2}) the system is also invariant under $s_0,s_2$ transformations :
\begin{equation}
s_2(q)= q+ \frac{\alpha_2}{p}, \;
s_2 (p)= p, \;
s_2(\alpha_1+\alpha_3) =  2\alpha_2+\alpha_1 +\alpha_3, \; 
s_2(\alpha_2)=-  \alpha_2 
\label{s2def}
\end{equation}
and
\begin{equation}
s_0(q)= q+ \frac{1-\alpha_2-\alpha_1-\alpha_3}{p+r_0 z}, \;
s_0 (p)= p, \;
s_0(\alpha_1+\alpha_3) =2-2 \alpha_2-\alpha_1-\alpha_3,  \;
s_0(\alpha_2)=  \alpha_2  \,.
\label{s0def}
\end{equation}

Together, these transformations satisfy the following relations :
\begin{equation} s_i^2=1=\pi_i^2, \; \pi^2 \pi_i \pi^2 =\pi_{i+2}, \;
 \pi^2 s_i \pi^2 =s_{i+2}, \; i=0,2,
\label{prop1}
\end{equation}
for \[ \pi^2(q)=-q, \; \pi^2(p)=-p - r_0 z, \; \pi^2(\alpha_i)=
\alpha_{i+2}, \; \pi^2(\epsilon_0)=-\epsilon_0
\]
as well as the commutation relations:
\begin{equation}   s_i s_{i+2}=s_{i+2}  s_i, \; \pi_i \pi_{i+2}=\pi_{i+2}  \pi_i, 
\;  \pi_i s_{i+2}=s_{i+2}\pi_i, \; i=0,2,
\label{prop2}
\end{equation}
that define the extended affine Weyl 
group $W[ s_0,s_2,\pi_0, \pi_2, \pi^2]$ as  established previously in
\cite{AGZ2016} (see equations (5.9) and (5.10) there).

One expects that this extended affine Weyl symmetry should define the 
model uniquely. The question is therefore if all these models
labeled by $J$ 
are really not equivalent to each other. 
To explore this question  we will cast the above transformations in a more  standard form by  first performing a
canonical transformation : 
\begin{equation*}
q \to {\tilde q} =q/z^{-(1+J)/2}, \quad
p \to {\tilde p} =pz^{-(1+J)/2}, 
\end{equation*}
with the Hamiltonian system of equations
\begin{align}
z {\tilde q}_z &= {\tilde q}^2 \left(2 {\tilde p}+r_0 z^{-(1+J)/2} \right)- 
\left((J-1)/2+\alpha_1+\alpha_3 \right) {\tilde q}
+\epsilon_0 r_0 z^{(1-J)/2} \nonumber \\
z {\tilde p}_z &=   {\tilde p} \left({\tilde p}+ r_0 z^{-(1+J)/2}  \right)
\left(- 2{\tilde q} \right) + 
((J-1)/2+ \alpha_1+\alpha_3) {\tilde p} - 
\alpha_2  r_0 z^{(1-J)/2} 
\label{tiqzpz}
\end{align}
that leads to simplified symmetry transformations by absorbing  factors 
like $z^{-(1+J)/2} $ appearing in e.g. (\ref{pi0}):
\begin{equation*}
\pi_0({\tilde q})= -\frac{\epsilon_0 }{{\tilde q}}, \quad  
\pi_0 ({\tilde p})= \frac{1}{ \epsilon_0 } \left( {\tilde q}^2 {\tilde p}+\alpha_2
{\tilde q} \right)
\end{equation*}

Furthermore for $J \ne 1$ we are able to define new variables $W,F$ as
\[
W ={\tilde q}/\sqrt{(1-J)/2},  \quad \quad 
F={\tilde p}/\sqrt{(1-J)/2}.\nonumber 
\]
The above transformation is not canonical, however  introducing
\[
\xi = z^{(1-J)/2}, \quad \text{ for}\;  J
\ne 1
\]
we can rewrite the corresponding equations as a Hamiltonian system
\begin{align}
\xi W_\xi &= \frac{\partial H_1}{\partial F}=W^2 \left(2F+{\hat r}_0 \xi \right)- 
\left({\hat \alpha}_1+{\hat \alpha}_3 \right) W
+{\hat \epsilon}_0 {\hat r}_0 \xi \nonumber \\
\xi F_\xi &= - \frac{\partial H_1}{\partial W}=  F \left(F+ {\hat r}_0 \xi \right)
\left( - 2W \right) + 
({\hat \alpha}_1+{\hat \alpha}_3) F - 
{\hat \alpha}_2  {\hat r}_0 \xi 
\label{WzFz}
\end{align}
with new parameters:
\[
{\hat r}_0 =r_0/\sqrt{(1-J)/2}, \; {\hat \epsilon}_0 =\epsilon_0/((1-J)/2), \;
{\hat \alpha}_2 =\alpha_2/((1-J)/2), \;
{\hat \alpha}_1+{\hat \alpha}_3= (C+\alpha_1+\alpha_3)/((1-J)/2),
\]
 with respect to the  new Hamiltonian:
\[
H_1= W^2F^2+W^2F {\hat r}_0 \xi - ({\hat \alpha}_1+{\hat \alpha}_3)WF+
{\hat \epsilon}_0 {\hat r}_0 \xi F +{\hat \alpha}_2  {\hat r}_0 \xi W\, .
\]
We note that with this association the following relation holds
\[ 
1-{\hat \alpha}_1-{\hat \alpha}_2-{\hat \alpha}_3= 
(1 -\alpha_1-\alpha_2 - \alpha_3)/((1-J)/2) 
\]
that shows that the system  is properly normalized for $J\ne 1$ with ${\hat
\alpha}_0 = \alpha_0/(1-J)/2$.

Therefore as long as $J\ne 1$  we were able to cast
the system  for $r_1 =  0$ and  general 
$J \neq 1$  into Hamilton equations (\ref{qzpz}) previously obtained 
for $J=-1, r_1=0$ with
$\alpha_i, i=0,1,2,3$  replaced by ${\hat \alpha_i}, i=0,1,2,3$. 
Thus, as as long as $J\ne 1$ the model obtained in $r_1 \to 0$ limit
is equivalent to P$_{III}$ model with  
an  extended affine Weyl 
group $W[ s_0,s_2,\pi_0, \pi_2, \pi^2]$ 
symmetry acting according to relations (\ref{pi0}), (\ref{pi2}), (\ref{s0def})
and
(\ref{s2def}) with $J=-1$.
In particular, we find by substituting $J=-1$ in
(\ref{pi0}), (\ref{pi2}), (\ref{s0def}) and (\ref{s2def})
that 
\[
v_1=    \frac{1}{2} \left(\alpha_0+\alpha_2\right), \qquad
v_2= \frac{1}{2}  \left( \alpha_0-\alpha_2 \right) \,,
\]
transform under $s_2,\pi^2,\pi_0$ as
\begin{equation}
\binom{v_1}{v_2} \stackrel{s_2}{\longrightarrow}
 \binom{v_2}{v_1}, \quad
 \binom{v_1}{v_2} \stackrel{\pi^2}{\longrightarrow}
 \binom{v_1}{-v_2} , \quad
\binom{v_1}{v_2} 
\stackrel{\pi_0}{\longrightarrow}
\binom{-1-v_2}{-1-v_1}
\label{pi0vv}
\end{equation}
One sees that actions of $\pi_0, s_2,\pi^2$ on parameters
$(v_1,v_2)$  realize a representation of the extended affine Weyl group for the root
system $C^{(1)}_2$ \cite{forrester,AGZ2016}. Consider namely a 2-dimensional vector space ${\mathbf V}$ consisting of vectors
${\mathbf v} = v_1 {\mathbf e_1} +v_2 {\mathbf e_2}$, with
${\mathbf e_1}, {\mathbf e_2}$ being a canonical basis of 
${\mathbf V}$. Define next a symmetric
bilinear form  $\langle \cdot | \cdot \rangle$ in ${\mathbf V}$
such that $\langle {\mathbf e_i} | {\mathbf e_j} \rangle=\delta_{ij}$.
Then according to \cite{okap3} vectors
\begin{equation}
{\mathbf a_1}=  {\mathbf e_1} - {\mathbf e_2}, \;\;
{\mathbf a_2}=  {\mathbf e_2} 
\label{B2roots}
\end{equation}
are the fundamental roots of the 
$C_2$ root system and
\begin{equation}
{\mathbf a_0}=  {\mathbf e_1} + {\mathbf e_2}
\label{B2highroot}
\end{equation}
is its highest root. Geometrically, the transformations $s_2, \pi^2$ are reflections
in the hyperplane perpendicular to vectors ${\mathbf a_i},i=1,2$
and the transformation $\pi_0$
corresponds to reflections in the hyperplane $\{ {\mathbf v}: 
\langle {\mathbf a_0} | {\mathbf v} \rangle=-1\}$ \cite{AGZ2016}.

As one can see from (\ref{pi0}), (\ref{pi2}) the transformation $\pi_0$ 
for the special value of  $J=1$  transforms $\alpha_i$ exactly as 
$s_0$ and $\pi_0 (v_1)=-v_2, \pi_0 (v_2)=-v_1$ no longer involves
reflection in the hyperplane perpendicular to the highest root.
Thus actions of these
transformations do not coincide in this case with an extended affine Weyl
symmetry  within this geometric interpretation. 

We now turn our
attention to the remaining case of reduction of 
the P$_{III-V}$ model for $r_1=0$ and $J=1$.


\subsection{\bf {The $\mathbf{r_1 \to 0}$ limit when  $\mathbf{J=1}$}}

We now consider separately the case $J=1$ when $r_1=0$. Inserting $J=1$ into
equations (\ref{tiqzpz}) and defining
\[ x =\ln z, \qquad w= {\tilde q}, \qquad f = {\tilde p}
\]
we obtain
\begin{equation}
\begin{split}
 w_x &= w^2 \left(2f+r_0 \right)- 
\left(\alpha_1+ \alpha_3-1 \right) w
+\epsilon_0 r_0 \\
f_x &=   f \left(f+ r_0 \right)
\left( - 2 w \right) + 
(\alpha_1+\alpha_3-1) f - 
\alpha_2  r_0
\end{split}
\label{wfz}
\end{equation}
that originate from a Hamiltonian
\begin{equation}
H= f^2w^2+w^2fr_0+(\alpha_1+\alpha_3-1) f w+ \epsilon_0 r_0 f+\alpha_2
r_0 w
\label{Hfw}
\end{equation}
The second order equation for $w$ is given by:
\[
w_{xx}= \frac{w_x^2}{w}+w^3r_0^2+w^2r_0( \alpha_2 -\alpha_0)
-\epsilon_0 r_0 (\alpha_0+\alpha_2)-\frac{\epsilon_0^2 r_0^2 }{w}
\]
and agrees with equation I$_{12}$ of Ince as reproduced in
(\ref{Ince:12}) in 
\ref{sec:ince}.

The second order equation for $f$ written in terms of $y$ such that
\[ f= -\frac{r_0 y}{y-1}
\]
leads to Ince's equation XXXVIII  (\ref{Ince:38}) 
with $\mathcal{A}=(1-\alpha_1-\alpha_2-\alpha_3)^2/2$, $\mathcal{B}
=-\alpha_2^2/{2}$, $\mathcal{C}=-2 \epsilon_0 r_0^2$
and  $\mathcal{D}=0$ and thus the equation obtained in this limit is
only an incomplete version of Ince's 38-th equation  (\ref{Ince:38}). 

\section{The $\mathbf{r_0 \to 0}$ limit of  P$_{\mathbf{III-V}}$ model} 
Setting $r_0 \to 0$ in equation 
(\ref{qzz}) yields
\begin{equation} 
q_{zz}  = - \frac{q_z}{z}+ \left( \frac{1}{2q}+\frac{1}{2(q-r_1)} \right)
q_z^2 +\frac{\alpha_1^2 r_1 (q-r_1)}{2 z^2 q}-
\frac{\alpha_3^2 r_1 q}{2 z^2 (q-r_1)}
-2 \epsilon_1 r_1 z^{J-1} q (q-r_1)
\label{qzzr00}
\end{equation}
For $J=0$ 
one recognizes in the above equation for $y = (q-r_1)/q$ 
the Painlev\'e V equation (\ref{P5})
with the 
parameter $\mathcal{D}=0$. For $\mathcal{D}=0$ the Painlev\'e V equation 
is known to be equivalent to
the Painlev\'e III equation \cite{gromak-book}.

Applying automorphism $\pi$   (\ref{pipbarq})
one transforms the symmetry transformations $\pi_i,s_i, i=0,2$
to symmetry transformations $\pi_i i=1,3$ :
\begin{equation}
\pi_1(p)= \frac{\epsilon_1 z^{J+1}}{p}, \;
\pi_1 (q)= \frac{ z^{-J-1}}{\epsilon_1 } \left(- p^2q+\alpha_1 p
\right), \;
\pi_1 (\alpha_1)=  \alpha_1 , \;
\pi_1(\alpha_3)= J+1-\alpha_3
\label{pi1}
\end{equation}
\begin{equation}
\pi_3(p)= -\frac{\epsilon_1 z^{J+1}}{p}, \;
\pi_3 (q)= \frac{ z^{_J-1}}{\epsilon_1 } \left( p^2(q-r_1)-
\alpha_3 p \right)+r_1, \;
\pi_3 (\alpha_1)=  J+1-\alpha_1 , \;
\pi_3(\alpha_3)= \alpha_3\, ,
\label{pi3}
\end{equation}
that together with transformations $s_1=\pi s_0 \pi$ and $s_3=\pi s_2 \pi$
keep invariant  equations 
\begin{equation}
\begin{split}
z q_z &= q \left(q-r_1 \right) 2p- 
\left(\alpha_1+\alpha_3 \right) q
+\alpha_1 r_1  \\
z p_z &=   p^2
\left(r_1 - 2q \right) + 
(\alpha_1+\alpha_3) p - 
\epsilon_1 r_1 z^{J+1}
\end{split}
\label{qzpzr0}
\end{equation}
obtained from (\ref{qzpz}) in the limit $r_0 \to 0$. 
For $J \ne -1$ the transformation 
\begin{equation*}
q \to q/z^{-(1+J)/2}/\sqrt{-(1+J)/2}=F, \quad
p \to =pz^{-(1+J)/2}/\sqrt{{-(1+J)/2}}=W,
\end{equation*}
followed by a change of variable $z \to \xi=z^{(1+J)/2}$ leads to equations:
\begin{equation}
\begin{split}
\xi W_\xi &=   W^2 (2 F- {\hat r}_1 \xi)
-({\hat \alpha}_1+{\hat \alpha}_3) W +
{\hat \epsilon}_1  {\hat r}_1 \xi \\
\xi F_\xi&=  F (F- {\hat r}_1 \xi) (-2 W) + 
\left({\hat \alpha}_1+{\hat \alpha}_3 \right) F
-{\hat \alpha}_1 {\hat r}_1 \xi
\end{split}
\label{FWr00}
\end{equation}
where ${\hat r}_1 = r_1/\sqrt{-(1+J)/2}, {\hat \epsilon}_1=
\epsilon_1/\sqrt{-(1+J)/2}, {\hat \alpha}_i= \alpha_i/\sqrt{-(1+J)/2}$.
One obtains from (\ref{FWr00}) the following second order equation for
$W$:
\[
W_{\xi \xi}=  \frac{W_\xi^2}{W}- \frac{W_\xi}{\xi}+ W^3 {\hat r}_1^2+W^2
{\hat r}_1 \frac{{\hat \alpha}_1-{\hat \alpha}_3+1}{\xi}+\frac{{\hat \epsilon}_1^2 {\hat r}_1^2}{W}
-\frac{{\hat r}_1}{\xi} {\hat 
\epsilon}_1 ({\hat \alpha}_1-{\hat \alpha}_3+1),
\]
which is Painlev\'e III equation (\ref{P3}). Thus we have obtained Painlev\'e III equation 
in $r_0 \to 0$ limit for any $J \ne -1$. This establishes another way
to understand an equivalence between Painlev\'e V equation (\ref{P5}) with
$\mathcal{D} =0$ and Painlev\'e III equation (\ref{P3}) realized in a
setting of Hamilton equations.

\subsection{\bf {The $\mathbf{r_0 \to 0}$ limit for $\mathbf{J=-1}$}} 
The transformations $\pi_i, i=1,3$ in (\ref{pi1}), (\ref{pi3})
for the special value of $J=-1$ transform $\alpha_i$ in the same
way as $s_1,s_3$ and it does not look in such case that actions of these
transformations on roots will  form  an extended affine Weyl
symmetry group. 
To investigate this further we set
$J=-1$ directly in (\ref{qzpzr0}) to obtain (for $x =\ln z$):
\[
\begin{split}
q_x&= q \left(q-r_1 \right) 2p- 
\left(\alpha_1+\alpha_3 \right) q
+\alpha_1 r_1 \, , \\
 p_x &=   p^2
\left(r_1 - 2q \right) + 
(\alpha_1+\alpha_3) p - 
\epsilon_1 r_1 \, .
\end{split}
\]
Let us set  as before $q=w,p=f$ and note that the Hamiltonian that
reproduces the above equation is given by :
\begin{equation}
H= f^2w^2-wf^2 r_1-(\alpha_1+\alpha_3) f w+ \epsilon_1 r_1 w+\alpha_1
r_1 f\,.
\label{HfwC0}
\end{equation}
Note that the major difference from (\ref{Hfw}) is the term $w f^2$ instead
for $w^2f$.

For the quantity $f=p$ we find from the above equations 
a second order equation: 
\[
f_{xx}= \frac{f_x^2}{f}+f^3 r_1^2+f^2r_1(-\alpha_1+\alpha_3)
+\epsilon_1 r_1 (\alpha_1+\alpha_3)-\frac{\epsilon_1^2 r_1^2 }{f}
\]
in which we again recognize the XII-th equation of Ince (\ref{Ince:12}).
Furthermore we derive:
\[
\begin{split}
w_{xx}&= \frac{w_x^2}{2}\left( \frac{1}{w}+\frac{1}{w-r_1}\right)-2
r_1 \epsilon_1 w^2 +\alpha_1^2 r_1\\
&+\frac{2 r_1^2 \epsilon_1 w^2}{w-r_1}-
w r_1 \frac{\alpha_1^2+\alpha_3^2+4 r_1^2 \epsilon_1}{2(w-r_1)}
+ \frac{r_1^3 \alpha_1^2}{2(w-r_1)}
\end{split}
\]
Defining $y$ in terms of $w$ as
\[
y= \frac{w}{w-r_1} \quad \text{or}  \quad w = \frac{r_1y}{y-1}
\]
one obtains 
a special case of Ince's equation I$_{49}$ (\ref{Ince:49}) listed in 
\ref{sec:ince} 
with the parameters $\mathcal{A}=1,
\mathcal{B}=\alpha_3^2/2,\mathcal{C}=-\alpha_1^2/2$
and $\mathcal{D}+\mathcal{E}=2 r_1^2 \epsilon_1$.

Note that Ince's equation I$_{38}$ (\ref{Ince:38}) with $\mathcal{D}=0$ can be rewritten as Ince's 
equation 49 (\ref{Ince:49}) with $\mathcal{A}=1$ and vice versa.
\section{Discussion}
One of main lessons derived from the above exercises of reducing 
P$_{III-V}$ is that the Hamilton functions of the type
\begin{equation}
H= f^2w^2+\kappa w f^2 +\beta f w+ \gamma f+\delta w,
\;\;\;\text{or}\;\;\;
H= f^2w^2+\kappa w^2 f +\beta f w+ \gamma f+\delta w
\label{hamkappa}
\end{equation}
will lead to I$_{12}$ equation and will be invariant under
the symmetry generators that satisfy the Coxeter group relations 
(\ref{prop1}), (\ref{prop2}).
Let us illustrate this using the first of Hamiltonians in (\ref{hamkappa}).
The corresponding Hamilton equations :
\begin{equation}
 w_x = 2 w^2 f+ 2 \kappa w f+
\beta w +\gamma, \;\quad
f_x =   -2 f^2 w- \kappa f^2 -\beta f - 
\delta \, .
\label{kappawfz1}
\end{equation}
lead to a second order equation for $f$:
\[
f_{xx}= \frac{f_x^2}{f} -\delta \beta
+f^2 (\kappa \beta -2 \gamma) +f^3 \kappa^2 -\delta^2/f
\]
which is I$_{12}$ equation (\ref{Ince:12}) from Ince's list.

Eqs. (\ref{kappawfz1}) are invariant under $s_2$ transformations:
\begin{equation}
s_2(f)= f+ \frac{\gamma}{\kappa w}, \;\,
s_2 (w)= w, \;\, s_2(\beta) =  \beta - 2 \gamma/\kappa, \;\,
s_2 (\kappa)=\kappa, \; \,
s_2(\delta)=  \delta ,  \; \, s_2 (\gamma)= -\gamma\, ,
\label{s2def1}
\end{equation}
and $\pi_0$ transformations :
\begin{equation}
\pi_0(f)= -\frac{\delta}{\kappa f}, \;\,
\pi_0 (w)= \frac{ \kappa}{\delta } 
\left( f^2w+\gamma f /\kappa\right),  \;\, 
\pi_0 (\beta) =  -\beta+2 \frac{\gamma}{\kappa},  \; \,
\pi_0 (\delta)=  \delta, \;\,
\pi_0(\gamma)= \gamma\, ,
\label{pi01}
\end{equation}
with $\pi_0(\kappa)=\kappa$ as well as a version of $\pi^2$:
\[
\pi^2 (f)=-f, \;\; \pi^2 (w)=-w-\kappa, \;\; 
\pi^2 (\gamma)= -\gamma +\kappa \beta
,\;\, \pi^2 (\delta)=-\delta, \; \pi^2 (\kappa)=\kappa,\,
\pi^2(\beta)=\beta \, .
\]
As in (\ref{prop1}) we can now define $s_0, \pi_2$ as 
$ \pi^2 \pi_0 \pi^2 =\pi_{2}, \;
 \pi^2 s_2 \pi^2 =s_{0}$ and obtain the Coxeter relations (\ref{prop2}).
As observed above the resulting  symmetry $W [s_0,s_2,\pi_0,\pi_2,\pi^2]$ 
can not be given the extended affine Weyl group interpretation that 
holds for $W [s_0,s_2,\pi_0,\pi_2,\pi^2]$ structure in the setting of
P$_{III}$ equation. Comparison of actions of $s_2$ and $\pi_0$ on 
parameters $\beta, \gamma, \delta,\kappa$ in equations (\ref{s2def1}) and
(\ref{pi01}) indeed reveals identical behavior (up to the sign) of those
two transformations, which in the discussion below (\ref{B2highroot}) was
recognized as a reason for why  the geometric interpretation of $W [s_0,s_2,\pi_0,\pi_2,\pi^2]$ 
as an extended affine Weyl group did not extend to the case of
symmetry of I$_{12}$ equation.

The remaining questions of how to complete Hamiltonian structures 
seen in this paper in such a way as to obtain full equations I$_{38}$, I$_{49}$ 
and what are the symmetries governing I$_{38}$, I$_{49}$  models
will be addressed in a paper in preparation \cite{Alves-etal}.
\subsection*{Acknowledgements}
  JFG and AHZ thank CNPq for financial support. VCCA thanks
  grant 2016/22122-9, S\~ao Paulo Research Foundation (FAPESP) for financial support.

\appendix
\section{Selected Equations from Ince's List}
\label{sec:ince}
Here we list the three equations, $I_{12}, I_{38}$ and $I_{49}$, from Ince's list, and two Painlev\'e equations that are subject of our discussion:
\begin{align}
I_{12}&:\,y_{xx}=  \frac{y_x^2}{y}+\mathcal{A} y^3+ \mathcal{B} y^2
+\mathcal{C}+\frac{\mathcal{D} }{y} 
\label{Ince:12}\\
I_{38} &:\, y_{xx} = \left( \frac{1}{2y}+\frac{1}{y-1} \right) y_x^2  
+ y(y-1)
 \left( {\cal A}(y-1) 
+{\cal B}  \frac{y-1}{y^2}+\frac{{\cal C} }{y-1}
+  \frac{{\cal D}}{(y-1)^2} \right)
\label{Ince:38} \\
I_{49} &:\,y_{xx}=\left(
\frac{1}{y}+\frac{1}{y-1}+\frac{1}{y-\mathcal{A}}\right)  \frac{y_x^2}{2}
+ y (y-1)(y-\mathcal{A})
\left( \mathcal{B}+ \frac{\mathcal{C}}{y^2}+\frac{\mathcal{D}}{(y-1)^2}
+\frac{\mathcal{E}}{(y-\mathcal{A})^2} \right)
\label{Ince:49}\\
P_{III} &:\,y_{zz}  = - \frac1z y_z+  \frac{y_z^2 }{y}
+{\cal A} \frac{y^2 }{z} +{\cal C} y^3
+ \frac{{\cal B}}{ z}+
\frac{{\cal D}}{y}
\label{P3}\\
P_{V} &:\,y_{zz} (z)=\left(\frac{1}{y-1}+\frac{1}{2 y}\right)
y_z^2-\frac{y_z}{z}+\frac{(y-1)^2 \left(\mathcal{A} 
	y+\frac{\mathcal{B} }{y}\right)}{z^2}+\frac{\mathcal{C} y}{z}+\frac{\mathcal{D} y
	(1+y)}{y-1}
\label{P5}
\end{align}

\end{document}